\documentstyle[11pt,newpasp,epsf,twoside]{article}

\def\edcomment#1{\iffalse\marginpar{\raggedright\sl#1\/}\else\relax\fi}
\reversemarginpar

\begin{document}

\title{Dynamics of Capture in the Restricted Three-Body Problem}
\author{Sergey A. Astakhov}
\affil{Department of Chemistry \& Biochemistry, Utah State University, Logan, Utah 84322-0300, USA}
\author{Andrew D. Burbanks, Stephen Wiggins}
\affil{School of Mathematics, University of Bristol BS8 1TW, UK}
\author{David Farrelly}
\affil{Department of Chemistry \& Biochemistry, Utah State University, Logan, Utah 84322-0300, USA}

\begin{abstract}
We propose a new dynamical model for capture
of irregular moons which identifies chaos as the 
essential feature responsible
for initial temporary gravitational trapping within a planet's Hill sphere. 
The key point is that incoming potential satellites get trapped in chaotic orbits
close to ``sticky'' KAM tori in the neighbourhood of the planet, possibly 
for very long times, so that the chaotic layer largely dictates 
the final orbital properties of captured moons.
\end{abstract}

\section{Introduction}
The often puzzling properties of the irregular satellites of the giant 
planets -- most of which have been discovered during the last six years
 (see Gladman et al. 2001; Hamilton 2003; Sheppard \& Jewitt 2003; references therein
 and IAU Circular 8193) -- provide a window into conditions in the early 
Solar System.

The general mechanism by which the irregular satellites were captured
is thought to involve the following steps (Heppenheimer \& Porco 1977; Pollack et al. 1979; Murison 1989; 
Peale 1999; Gladman et al. 2001); 
(i) temporary trapping close to the planet in a region roughly demarked by the
 Lagrange points $L_1$ and $L_2$;
(ii) gradual energy loss through dissipation (e.g., gas drag or planetary growth) which translates 
temporary trapping into permanent capture; and (iii) 
possible fragmentation due to collisions at much later times.
The hypothesis that the observed clustering among populations
of irregular moons may be a result of fragmentation (Pollack et al. 1979; Gladman et al. 2001)
contradicts, however, the fact that the orbits of known irregulars
are clustered in inclination but not necessarily in eccentricity or other orbital elements (Nesvorny et al. 2003). 
Although there have been extensive studies of how the systems of irregular
sattelites have formed (see also Henon 1970; Colombo \& Franklin 1971; Huang \& Innanen 1983;
Saha \& Tremaine 1993; Gor'kavyi \& Taidakova 1995; Marzani \& Scholl 1998; 
Namouni 1999; Viera Neto \& Winter 2001; Winter \& Viera Neto 2001; 
Carruba et al. 2002; Carruba et al. 2003; Nesvorny et al. 2002; Nesvorny et al 2003; Winter et al. 2003) a coherent {\it dynamical} 
picture of capture has not emerged; e.g., it has been widely held that the propensity for retrograde motion 
among Jupiter's irregulars is simply due to the well known enhanced stability of retrograde 
orbits with large semimajor axes $a$ (Nesvorny et al. 2003). Gladman et al. (2001) have called this, 
and the alternative ``pull-down'' (Heppenheimer \& Porco 1977) capture
mechanism, into question based on the following observation: while the
bulk of Jupiter's irregular moons are retrograde and lie distant from 
the planet, Saturn's cortege contains a more even mix of prograde and 
retrograde moons even though they have similarly large semimajor axes $a$ when 
expressed in planetary radii. Here, we study capture in the circular 
restricted three-body problem (CRTBP) in two and three-dimensions (3D) 
taking the Sun-Jupiter-moon system as the specific example: 
the dynamical picture that emerges in the Hill limit (Murray \& Dermot 1999; Simo \& Stuchi 2000) is, however, rather 
similar for the other giant planets.

\section{The Hamiltonian}
In a coordinate system rotating with the mean motion of the primaries, 
but with origin transformed to the planet, the CRTBP Hamiltonian is given by

\[
H=E=\frac{1}{2}{\bf p}^{2}-(x\,p_{y}-y\,p_{x})-\frac{\mu }{\sqrt{x^{2}+y^{2}+z^{2}}}
\]
\begin{equation}
 -\frac{1-\mu }{\sqrt{(1+x)^{2}+y^{2}+z^{2}}}-(1-\mu)x + \alpha
\end{equation}

\noindent where $a$ is scaled to $1$, $\mu = m_{1}/(m_{1}+m_{2})$; $m_{1}$ and $m_{2}$ 
are the masses of the primaries and $\alpha$ is a collection of inessential constants retained for consistency in relating the
energy $E$ to the Jacobi constant $C_J = -2E$; ${\bf r}=(x,y,z)$ and ${\bf p}=(p_{x},p_{y},p_{z})$ are the coordinates 
and momenta of the potential satellite. This Hamiltonian is obtained from the standard CRTBP Hamiltonian (Murray \& Dermot 1999)
by the canonical transformation $x\rightarrow x{^{\prime }}+(1-\mu)$, $p_y \rightarrow p_{y}^{\prime }+(1-\mu)$ 
(and dropping the primes). Angular momentum, ${\bf h}=(h_x,h_y,h_z)$ where 
$h_z = x\,p_{y}-y\,p_{x}$, is now defined with respect to the planet as is natural for a study of capture.

\begin{figure}[ht]
\begin{center}
\epsfbox{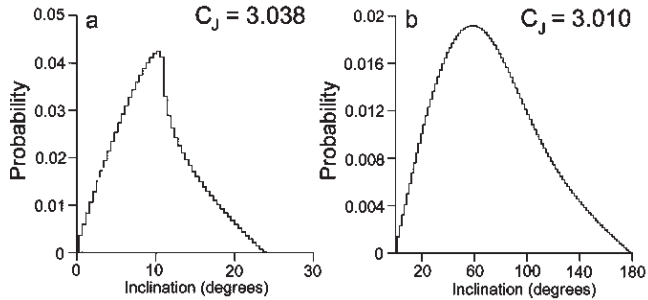}
\end{center}
\caption{Histograms of inclination distribution for $10^8$ test particles originating at the Hill sphere for indicated values
of energy (Jacobi constant).}
\end{figure}

\section{Simulations}
Figures 1 shows computed orbital inclination distributions ($I
= \arccos(h_z/h)$) for a flux of $10^8$ test particles as they pass through the Hill 
sphere (radius $R_H = a(\mu/3)^{1/3}$, Murray \& Dermot (1999)) at two energies. The key observation is that at low
energy (Figure 1a) only {\it prograde} orbits can enter (or exit) the capture zone between 
Lagrange points $L_1$ and $L_2$ whereas at higher energies (Figure 1b) the distribution shifts to
include both senses of $h_z$.  This is because not all parts of the Hill sphere are 
energetically accessible at low energies. Figure 1 thus suggests that 
the statistics of capture might be expected to depend on initial $C_J$ and ${\bf h}$. 

To investigate this we consider first the structure of phase space in
the planar limit ($z = p_z = 0$). Figure 2 displays 
a series of Poincar\'e surfaces of section (SOS) for randomly chosen initial
conditions inside the Hill radius $R_H$. The hypersurface is 
the $x-y$ plane with units rescaled to $R_H = 1$ and points colored according to the sign of angular
momentum (grey, retrograde $h_z < 0$; black, prograde $h_z > 0$) as they intersect the surface with $p_x = 0$ and $dy/dt > 0$.

\begin{figure}[ht]
\begin{center}
\epsfbox{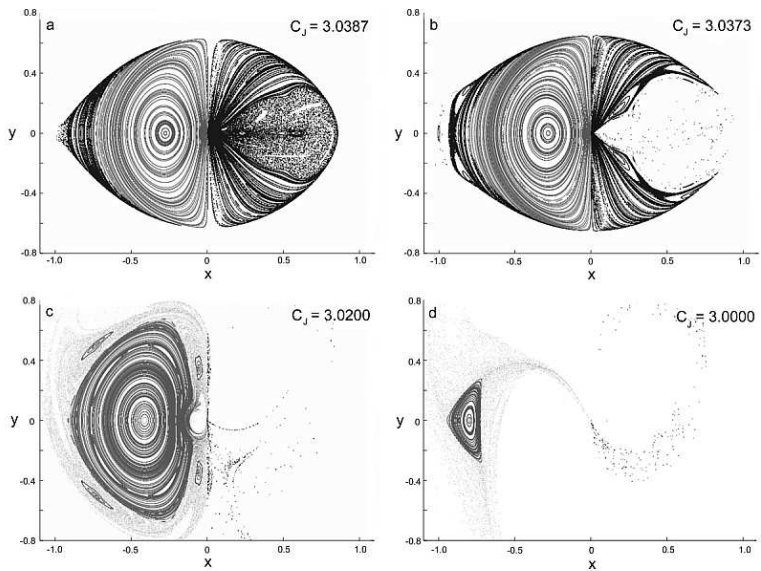}
\end{center}
\caption{Poincare surfaces of section for Sun-Jupiter-satellite 2D CRTBP with increasing energy.}
\end{figure}

With increasing energy a coherent dynamical picture emerges; in Fig. 2a the 
prograde orbits exhibit regions of strongly chaotic motion whereas 
{\it all} the retrograde orbits are regular (quasiperiodic). This 
forces incoming prograde orbits to remain prograde because they cannot 
penetrate the KAM regions in Figure 2a. Although KAM tori in 3D cannot ``block'' trajectories, 
if these regions are near-integrable then orbits can only enter by Arnold
diffusion which, by the Nekhoroshev theorem (Nekhoroshev 1977), is expected to occur
exponentially slowly. So under these conditions the 2D picture
should hold, in practice, also in 3D. Our simulations
indicate that it does. After the gateway at $L_2$ has opened in Figure 2b 
the chaotic ``sea'' of prograde orbits visible in Figure 2a rapidly disappears except for 
a thin residual front of chaos which sticks to the KAM tori, separating them from the
growing basin of direct scattering. As energy increases further this 
front moves from prograde to retrograde motion while the tori
steadily erode. KAM tori are ``sticky'' and chaotic 
orbits near them can appear locally near-integrable, i.e.,
they are trapped in almost regular orbits for very long times (Perry \& Wiggins 1994). 
Note especially that the KAM tori in Figures 2c and
2d exist at energies well {\it above} $L_1$ and $L_2$. Permanent capture happens 
if dissipation is sufficient to switch long lived chaotic orbits into 
KAM regions which means that chaotic orbits can be permanently 
captured, even above the saddle points by relatively weak dissipation.

The SOS also reveal that large distance from the planet need not
imply retrograde motion: e.g., the orbits visible inside two KAM islands centered at $x > 0, y =0$ in Figure 2a
are large, almost circular, periodic {\it prograde} orbits but whose centers 
are displaced from the origin. At higher energies capture (through 
dissipation) is into retrograde KAM surfaces nested around the
circular, retrograde orbit ($x < 0, y =0$) (Henon 1970; 
Winter \& Viera Neto 2001), and which remains almost perfectly centered on the planet.

In 3D, initial conditions in {\bf r} were chosen uniformly and
randomly on the Hill sphere with random velocities; The Jacobi
constant was also chosen randomly and uniformly $C_J \in
(2.995,C_J^{L_1})$. Trajectories were integrated until they exited the
Hill sphere, came within 2 planetary radii of the origin (Carruba et al. 2002) or
survived for a predetermined cutoff time $t_{cut}$.
Figure 3a shows a clear trend from prograde to retrograde capture with increasing energy.  
Three main islands stand out in the archipelago visible in Figure 3a; the prograde
(low inclination) island shrinks noticeably with increasing $t_{cut}$
reflecting the lower probability of prograde capture.

\begin{figure}[ht]
\begin{center}
\epsfbox{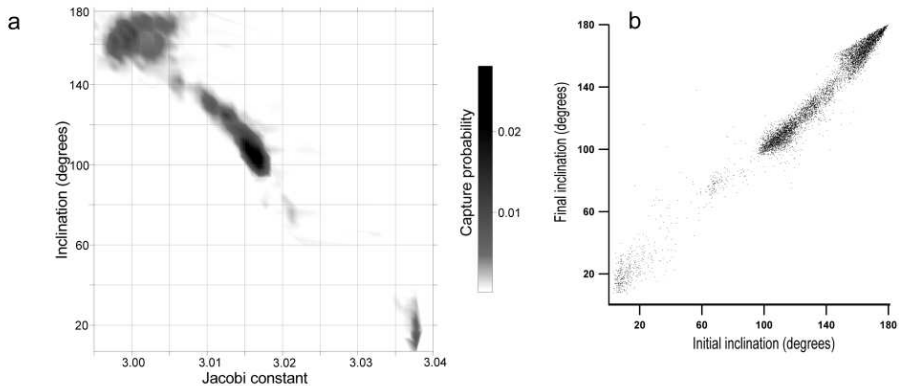}
\end{center}
\caption{(a) The normalized capture probability distribution from Monte Carlo simulations for 80 million 
test particles and $t_{cut} = 20000$ years (Jupiter), and (b) correlation between initial and final
inclination of the test particles trapped in the capture zone.}
\end{figure}

The most noticeable feature is the large island at $I \approx 100^\circ$
whose stability is related to the Kozai resonance centered at $I =
90^\circ$ (Kozai 1962; Innanen et al. 1997; Carruba et al. 2002; Carruba et al. 2003; Nesvorny et al. 2003). 
Unlike in 2D direct injection into KAM regions is possible but in near-integrable regions occurs 
exponentially slowly which seems to be why incoming particles are excluded
from the center of the resonance itself. Likely these particles are
trapped in a chaotic separatrix layer of the Kozai resonance.  
Both the ``Kozai island'' and the smaller, very high energy
island of retrograde motion in Figure 3a are stable for extremely long
times. Figure 3b demonstrates a very strong correlation between final and initial
angular momentum (i.e., ``inclination memory'', Astakhov et al. 2003) for long lived orbits which is
consistent with the picture in 2D. However, these are {\it post
facto} correlations since knowledge of initial inclination and total
energy is generally not a good predictor of lifetime.

We have confirmed numerically that these distributions are similar for the
other giant planets; are robust in the presence of gas drag of different forms.
The specific details of capture depend sensitively on how this dynamical 
mechanism intersects local environment around the planet. One of the
factors is the part of the Hill sphere occupied by the massive regular
moons. This observation is important, because prograde orbits penetrate
much deeper towards the planet than do most of the retrograde orbits
(see Figures 2b, 2d and also Simo \& Stuchi 2000). Therefore, to be
permanently captured, prograde satellites must survive close encounters
or collisions with ``influential'' regular moons. In our Monte Carlo simulations 
with dissipation (Astakhov et al. 2003) we eliminated test particles
that crossed the orbits of Titan at Saturn and Callisto at Jupiter.
These simulations indicate that Saturn/Titan in tandem have a clear
tendency to capture a higher ratio of prograde to retrograde
moons as compared to Jupiter/Callisto. This is because Callisto's orbit 
represents a larger fraction of the Hill sphere than does Titan's orbit. Thus the relative scarcity of jovian 
prograde irregulars may be due to potential prograde satellites
having been swept away more efficiently by Jupiter's Galilean moons (Astakhov et al. 2003).

\mbox{}

\noindent {\bf Acknowledgements}
This work was funded by grants from the US National Science Foundation and Petroleum Research 
Fund to DF.

\end{document}